\documentclass{comjnl}

\usepackage{amsmath}
\usepackage{mathrsfs}
\usepackage{amsfonts,amssymb}

\usepackage{xspace}

\usepackage{threeparttable}
\usepackage{multirow}
\usepackage{booktabs}
\usepackage{graphicx}
\usepackage{subcaption}

\usepackage{hyperref}
\usepackage{floatrow}
\usepackage{cite}
\usepackage{array}
\usepackage{verbatim}
\usepackage{xcolor}
\usepackage{colortbl}

\newcommand{\simon}{\textsc{Simon}\xspace}
\newcommand{\Simon}{\textsc{Simon}\xspace}
\newcommand{\SIMON}{\textsc{Simon}\xspace}
\newcommand{\SIMONnospace}{\textsc{Simon}}

\newcommand{\Simeck}{\textsc{Simeck}\xspace}
\newcommand{\SIMECK}{\textsc{Simeck}\xspace}
\newcommand{\SIMECKnospace}{\textsc{Simeck}}

\newcommand{\speck}{\textsc{Speck}\xspace}
\newcommand{\Speck}{\textsc{Speck}\xspace}

\newcommand{\SPECKnospace}{\textsc{Speck}}

\definecolor{mygray}{gray}{.95}

\makeatletter
\newcommand\dlmu[2][4cm]{\hskip1pt\underline{\hb@xt@ #1{\hss#2\hss}}\hskip3pt}
\makeatother

\begin{document}

\title[Improved (Related-key) Differential-based Neural Distinguishers]{Improved (Related-key) Differential-based Neural Distinguishers for SIMON and SIMECK Block Ciphers}

\author{Jinyu Lu$^{1,2,3}$}
\author{Guoqiang Liu$^{1,2,4*}$}

\author{Bing Sun$^{1,2,3}$}

\author{Chao Li$^{1,2,3}$}

\author{Li Liu$^{5,6}$}

\affiliation{$^\emph{1}$College of Liberal Arts and Sciences, National University of Defense Technology, Hunan, Changsha 410073, China\\ 
	$^\emph{2}$Hunan Engineering Research Center of Commercial Cryptography Theory and Technology Innovation, Hunan, Changsha 410073, China\\ 
	$^\emph{3}$State Key Laboratory of Cryptology, P.O.Box 5159, Beijing 100878, China\\
	$^\emph{4}$State Key Laboratory of Information Security (Institute of Information Engineering, Chinese Academy of Sciences, Beijing 100093, China\\
	$^\emph{5}$College of Systems Engineering, National University of Defense Technology, Hunan, Changsha 410073, China\\
	$^\emph{6}$Center for Machine Vision and Signal Analysis, University of Oulu, Oulu 90570, Finland}

\email{liuguoqiang87@hotmail.com}

\shortauthors{J. Lu, G. Liu, B. Sun, C. Li, L. Liu}



\begin{abstract}
In CRYPTO 2019, Gohr made a pioneering attempt and successfully applied deep learning to the differential cryptanalysis against NSA block cipher \SPECKnospace32/64, achieving higher accuracy than the pure differential distinguishers. By its very nature, mining effective features in data plays a crucial role in data-driven deep learning. In this paper, in addition to considering the integrity of the information from the training data of the ciphertext pair, domain knowledge about the structure of differential cryptanalysis is also considered into the training process of deep learning to improve the performance. Meanwhile, taking the performance of the differential-neural distinguisher of \SIMONnospace32/64 as an entry point, we investigate the impact of input difference on the performance of the hybrid distinguishers to choose the proper input difference. Eventually, we improve the accuracy of the neural distinguishers of \SIMONnospace32/64, \SIMONnospace64/128, \SIMECKnospace32/64, and \SIMECKnospace64/128. We also obtain related-key differential-based neural distinguishers on round-reduced versions of \SIMONnospace32/64, \SIMONnospace64/128, \SIMECKnospace32/64, and \SIMECKnospace64/128 for the first time. 
\end{abstract}

\maketitle

\thispagestyle{empty}

\section{Introduction}
The security analysis of many cryptographic primitives (such as pseudo-random number generators, hash functions, etc.) is usually attributed to attacks on the underlying block ciphers. Various cryptanalytic methods have been proposed over the past few decades, including differential cryptanalysis~\cite{biham1991differential}, linear cryptanalysis~\cite{matsui1993linear}, integral cryptanalysis~\cite{knudsen2002integral}, zero-correlation linear cryptanalysis~\cite{bogdanov2014linear}, etc. A block cipher must be able to resist all known cryptanalysis to obtain a strong security statement. In recent years, solver-based automatic tools and dedicated heuristic search algorithms have been extensively adopted to improve the accuracy and efficiency in cryptanalysis of block ciphers, where the cryptanalytic models are often transformed into MILP problems~\cite{mouha2011differential, sun2014automatic}, SAT/SMT problems~\cite{mouha2013proof, kolbl2015observations} or CP problems~\cite{minier2014solving, gerault2016constraint}. Automatic search technology has improved the analysis ability of block ciphers. The improvement and development of these automatic search technologies provide an inexhaustible source of thought for the design and analysis of block ciphers. However, these search technologies do not extract any new features that are not available manually. Therefore, once optimal distinguishers are obtained, these automatic tools would exert less influence in improving attacks.

Recently, under the joint driven form of big data and the availability of computing hardware, deep learning~\cite{lecun2015deep, bengio2021deep} has made remarkable progress and spread over almost every field of science and technology. Some researchers explored the feasibility of applying machine learning to the field of cryptography. In ASIACRYPT 1991, Rivest~\cite{rivest1991cryptography} made preliminary explorations of the possible connection between cryptography and machine learning, and some researchers applied machine learning in side channel analysis successfully, such as~\cite{maghrebi2016breaking, hospodar2011machine}. However, few researchers focused on the application of machine learning to black box cryptanalysis, until the process of applying deep learning to black box cryptanalysis was accelerated by the remarkable work of Gohr~\cite{gohr2019improving}.

Deep learning algorithms can analyze data and learn effective patterns for predicting new samples. Based on this, Gohr trained a deep neural network using the labeled (labels 0 and 1) ciphertext pairs as training data, where the data with label 1 comes from the encrypted plaintext pair with fixed input difference, and the data with label 0 is a random number. The trained neural network then is used to distinguish between the real ciphertext pairs and random pairs. When his network is applied to \SPECKnospace32/64, higher accuracy than the classical differential (CD) is achieved. Although the number of rounds using his network has not yet surpassed the number of rounds achieved by the most advanced technology, the neural distinguisher (ND) under the same number of rounds uses some information that the CD has not tapped.

More importantly, a potent key recovery attack is created by combining NDs with CDs and highly selective key search strategies. In essence, the NDs are too short to be used in key recovery and must be prepended with CDs to get the hybrid distinguishers (HDs). Making the resulting HDs usable in a key recovery attack requires better NDs or prepended CDs. Researchers have provided solutions from various angles. Benamira \emph{et al.}~\cite{benamira2021deeper} analyzed and explained the inner workings of Gohr's neural network and enhanced the accuracy of the NDs by creating batches of ciphertext inputs instead of pairs. Bao \emph{et al.}~\cite{bao2022enhancing} enhanced the CD's neutral bits and trained better NDs by investigating different neural networks, enabling key recovery attacks for the 13-round \SPECKnospace32/64 and 16-round \SIMONnospace32/64.

\begin{figure*}
	\centering
	\includegraphics[scale=0.5]{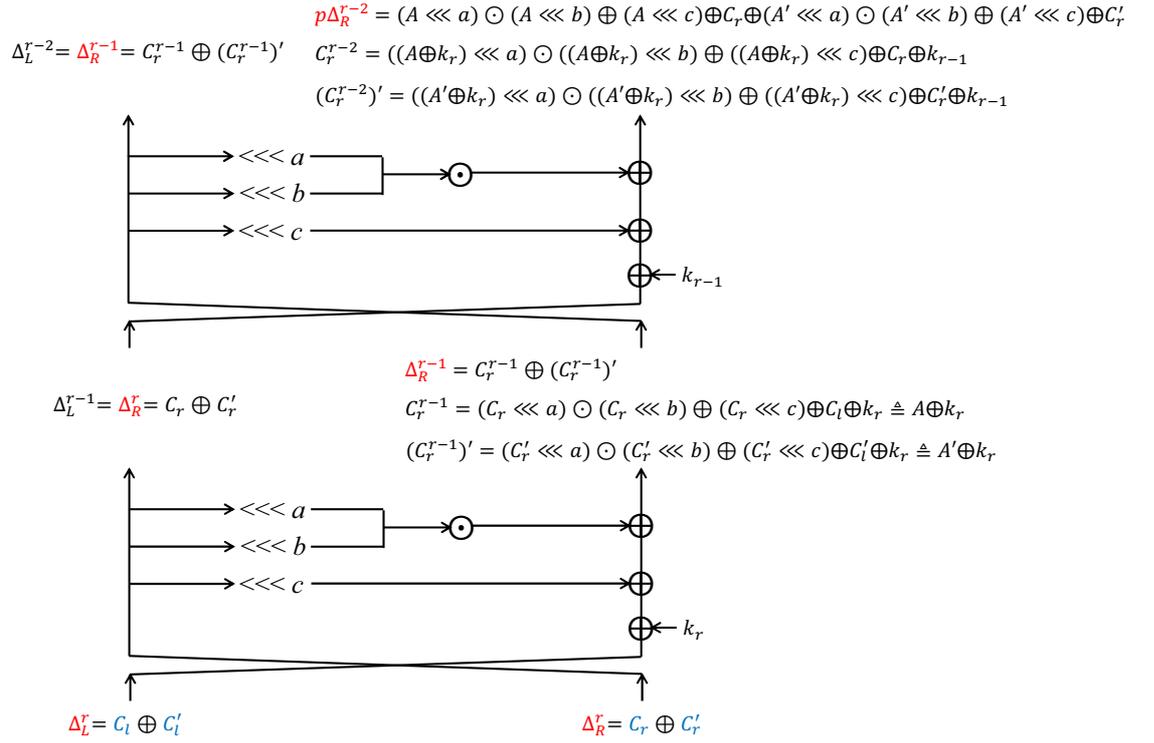}
	\caption{Notation of the data format.}\label{fig:data-format}
\end{figure*}

\paragraph{Our contribution:}
\begin{itemize}
	\item In this paper, we present (related-key) differential-based neural distinguishers on \Simon and \Simeck block ciphers. To better match our neural network and increase the accuracy of the neural distinguisher, we adopt the multiple ciphertext pairs (8 ciphertext pairs) to train the neural network fed with the data of form $(\Delta_L^{r}, \Delta_R^{r}, C_l, C_r, C_l' ,C_r', \Delta_R^{r-1}, p\Delta_R^{r-2})$. Fig.~\ref{fig:data-format} shows a schematic representation of these notations. Also, we employ the SE-ResNet network (Fig.~\ref{fig:network}) due to the success of ResNet on \Speck~\cite{gohr2019improving} and SENet on \Simon~\cite{bao2022enhancing}, as well as their superior performance on classification tasks.

	\item We notice that the choice of the ND or connecting difference  is critical to obtain the best hybrid distinguishers. Therefore, taking the performance of the differential-neural distinguisher of \SIMONnospace32/64 as an entry point, we investigate the impact of input difference on the performance of the hybrid distinguishers to choose the proper input difference. As a result, the input difference \textbf{\texttt{(0,$e_i$)}} is a good choice to obtain hybrid distinguishers for \simon-like ciphers.

	\item Eventually, we build neural distinguishers for \SIMONnospace32/64, \SIMONnospace64/128, \SIMECKnospace32/64 and \SIMECKnospace64/128. The results are shown in Table~\ref{tab:overview of results}, which shows that we improve the accuracy of the distinguishers. Meanwhile, we successfully construct the related-key neural distinguishers against \SIMONnospace32/64, \SIMONnospace64/128, \SIMECKnospace32/64 and \SIMECKnospace64/128 for the first time. 	
\end{itemize}

In this paper, the experiment is conducted by Python 3.6.10 in Ubuntu 18.04. The models are implemented by Tensorflow 2.5.0. The experiment uses a server with Intel(R) Xeon(R) Gold 6248 CPU *4 with 2.50GHz, 512GB RAM, and NVIDIA Tesla T4 16GB. The source code is available on Github\footnote{\url{https://github.com/JIN-smile/Improved-Related-key-Differential-based-Neural-Distinguishers}}.

\begin{table*}
	\centering
	\renewcommand\arraystretch{1.2}
	\newcommand{\tabincell}[2]{\begin{tabular}{@{}#1@{}}#2\end{tabular}}
	\caption{The comparison of (related-key) neural distinguishers attacks on \SIMONnospace32/64, \SIMONnospace64/128, \SIMECKnospace32/64, and \SIMECKnospace64/128 with 8 ciphertext pairs as a sample. ND: neural distinguisher, RKND: related-key neural distinguisher. TPR: True Positive Rate, TNR: True Negative Rate. \dag: For NDs fed with single ciphertext pairs, the combine-response distinguisher (CRD) obtained for the case of 8 ciphertext pairs. *: This neural distinguisher is trained using the staged training method.}
	\label{tab:overview of results}
	
	\scalebox{1}{
		
		\begin{threeparttable}
			
			\begin{tabular}{p{1.4cm}<{\centering}p{1.2cm}<{\centering}p{1cm}<{\centering}p{5cm}<{\centering}p{1.5cm}<{\centering}p{1.5cm}<{\centering}p{1.5cm}<{\centering}p{0.9cm}<{\centering}}
				\toprule
				Ciphers & \tabincell{c}{Attack\\ Model} & Round &  Input difference & Accuracy &TPR & TNR &Source \\
				\midrule
				\multirow{12}{*}{\tabincell{c}{\SIMONnospace\\32/64}} & \multirow{8}{*}{ND} & 9\dag & \texttt{(0x0,0x40)} & 0.8940 & 0.8728 & 0.9152 & ~\cite{bao2022enhancing} \\
				
			    &  & 9 & \texttt{(0x0,0x40)} &  0.9176 &  0.9052 &  0.9299 & ~\ref{sec:app-simon} \\
				\cline{3-8}

				&  & 10*\dag & \texttt{(0x0,0x40)} &  0.6865 &  0.6817 &  0.6912 & ~\cite{bao2022enhancing} \\
				&  & 10 & \texttt{(0x0,0x40)} & 0.6975 &  0.6662 &  0.7287  &  ~\ref{sec:app-simon} \\
				
				\cline{3-8}
				
				&  & 11*\dag & \texttt{(0x0,0x40)} &  0.5568 &  0.5419 &  0.5717 & ~\cite{bao2022enhancing} \\
				&  & 11 & \texttt{(0x0,0x40)} & 0.5609 &  0.5366 &  0.5852  &  ~\ref{sec:app-simon} \\
				
				\cline{3-8}
				
				&  & 12 & \texttt{(0x1,0x4)} &  0.5152 &  0.4799 &  0.5505 & \multirow{2}{*}{~\ref{sec:app-simon}} \\
				&  & 12* & \texttt{(0x0,0x40)} & 0.5142 &  0.5029 &  0.5254  & \\
				
				
				\cline{2-8}
				&  \multirow{4}{*}{RKND}& 10 & \texttt{(0x0,0x40),(0x0,0x0,0x0,0x40)}  & 1 &  1 &  1 & \multirow{4}{*}{~\ref{sec:app-simon}} \\
				&  & 11 & \texttt{(0x0,0x40),(0x0,0x0,0x0,0x40)}&  0.9604 &  0.9639 &  0.9569 &  \\
				&  & 12 & \texttt{(0x0,0x40),(0x0,0x0,0x0,0x40)}&  0.6477 &  0.6518 &  0.6435 &  \\
				&  & 13 & \texttt{(0x0,0x40),(0x0,0x0,0x0,0x40)}&  0.5262 &  0.5437 &  0.5081 &  \\
				\hline
				\hline
				
				
				\multirow{7}{*}{\tabincell{c}{\SIMECKnospace\\32/64}} & \multirow{4}{*}{ND} & 9 & \texttt{(0x0,0x40)} &  0.9952 & 0.9989 &  0.9914 & \multirow{4}{*}{~\ref{sec:app-simeck}}  \\
				
				&  & 10 & \texttt{(0x0,0x40)} &  0.7354 & 0.7207 &  0.7501 &  \\
				&  & 11 & \texttt{(0x0,0x40)} &  0.5646 & 0.5356 &  0.5936 &  \\
				
				\cline{3-7}
				
				&  & 12* & \texttt{(0x0,0x40)} &  0.5146 & 0.4770 &  0.5522 &  \\

				\cline{2-8}
				&  \multirow{3}{*}{RKND}& 13 & \texttt{(0x0,0x40),(0x0,0x0,0x0,0x40)}  &  0.9950 &  0.9990 &  0.9910 & \multirow{3}{*}{~\ref{sec:app-simeck}} \\
				&  & 14 & \texttt{(0x0,0x40),(0x0,0x0,0x0,0x40)} & 0.6679 &  0.6425 &  0.6933 &  \\
				&  & 15 & \texttt{(0x0,0x40),(0x0,0x0,0x0,0x40)} & 0.5467 &  0.5173 &  0.5762 &  \\
				\hline
				\hline
				
				
				\multirow{8}{*}{\tabincell{c}{\SIMONnospace\\64/128}} & \multirow{5}{*}{ND} & 11 & \texttt{(0x0,0x40)} & 0.9181  & 0.9045 & 0.9318 & \multirow{5}{*}{~\ref{sec:app-simon}} \\

				&  & 12 & \texttt{(0x0,0x40)} &  0.7117 &  0.6705 &  0.7530 &  \\
				
				&  & 13 & \texttt{(0x0,0x40)} &  0.5722 &  0.5230 &  0.6215 &  \\
				
				\cline{3-7}
				
				&  & 14 & \texttt{(0x0,0x40)} &  0.5148 &  0.4697 &  0.5600 &  \\
				
				&  & 14* & \texttt{(0x0,0x40)} &  0.5185 &  0.4663 &  0.5707 &  \\

				\cline{2-8}

				&  \multirow{3}{*}{RKND}& 12 & \texttt{(0x0,0x40),(0x0,0x0,0x0,0x40)}  & 0.9880  & 0.9894  & 0.9865  & \multirow{3}{*}{~\ref{sec:app-simon}}  \\
				
				& & 13 & \texttt{(0x0,0x40),(0x0,0x0,0x0,0x40)}  & 0.8398  & 0.8389   & 0.8408  &  \\
				
				& & 14 & \texttt{(0x0,0x40),(0x0,0x0,0x0,0x40)}  & 0.5788  & 0.5894   & 0.5682  &  \\

				\hline
				\hline
				
				
				\multirow{11}{*}{\tabincell{c}{\SIMECKnospace\\64/128}}& \multirow{6}{*}{ND} & 14 & \texttt{(0x0,0x40)} & 0.9142  & 0.8914  & 0.9371 &  \multirow{6}{*}{~\ref{sec:app-simeck}} \\
				& & 15 & \texttt{(0x0,0x40)} & 0.7663  &  0.6981 &  0.8345 &  \\
				&  & 16 & \texttt{(0x0,0x40)} &  0.6356 &  0.5245 &  0.7467 &  \\
				&  & 17 & \texttt{(0x0,0x40)} &  0.5577 &  0.4301 &  0.6853 &  \\
				
				\cline{3-7}
				&  & 18 & \texttt{(0x0,0x40)} &  0.5202 &  0.3917 &  0.6486 &  \\
				&  & 18* & \texttt{(0x0,0x40)} &  0.5218 &  0.3927 &  0.6510 &  \\

				\cline{2-8}
				&  \multirow{5}{*}{RKND}  & 18 & \texttt{(0x0,0x40),(0x0,0x0,0x0,0x40)}  &  0.9066 & 0.8837 &  0.9295 & \multirow{5}{*}{~\ref{sec:app-simeck}} \\
				& & 19 & \texttt{(0x0,0x40),(0x0,0x0,0x0,0x40)}  &  0.7558 &  0.6845 &  0.8270 & \\
				& & 20 &  \texttt{(0x0,0x40),(0x0,0x0,0x0,0x40)} & 0.6229 &  0.5104 &  0.7354 &  \\
				& & 21 &  \texttt{(0x0,0x40),(0x0,0x0,0x0,0x40)} & 0.5519 &  0.4248 &  0.6790 &  \\
				& & 22 &  \texttt{(0x0,0x40),(0x0,0x0,0x0,0x40)} & 0.5180 &  0.3906 &  0.6455 &  \\

				\bottomrule
			\end{tabular}
	\end{threeparttable}}
\end{table*}

\paragraph{Organization.}
Section~\ref{sec:preli} recalls \Simon-like ciphers, (related-key) differential cryptanalysis and CNN network. Section~\ref{sec:improve} introduces improved (related-key) differential-based neural distinguishers, including the batches of ciphertext pairs with new data format, and the network architecture. Section~\ref{sec:input difference} compares the performance of the hybrid distinguisher with different input difference. Section~\ref{sec:app-simon} gives the (related-key) differential-neural distinguishers for round-reduced \SIMONnospace32/64 and \SIMONnospace64/128. Section~\ref{sec:app-simeck} provides the (related-key) differential-neural distinguishers for round-reduced \SIMECKnospace32/64 and \SIMECKnospace64/128. Section~\ref{sec:conclusion} concludes this paper.

\section{Related works}\label{sec:preli}

\subsection{Notations}
Table~\ref{tab:notation} presents the notations used in this paper.

\begin{table}
	\scalebox{0.66}{
		\caption{The notations used throughout the paper}
		\label{tab:notation}
		\begin{tabular}{lp{9cm}}
			\toprule
			Notation & Description \\
			\midrule
			$x=(x_{n-1},\ldots,x_0)$ & Binary vector of $n$ bits; $x_i$ is the bit in position $i$ with $x_0$ the least significant one. \\
			$x \odot y$ & Bitwise AND between $x$ and $y$. \\
			$x \oplus y$ & Bitwise XOR between $x$ and $y$. \\
			$x \ \Vert \ y$ & Concatenation of $x$ and $y$. \\
			$x \lll \gamma,S^{\gamma}(x)$ & Circular left shift of $x$ by $\gamma$ bits. \\
			$x\ggg \gamma,S^{-\gamma}(x)$ & Circular right shift of $x$ by $\gamma$ bits. \\
			$(P_l, P_r, P_l', P_r')$ &  A set of plaintext pairs with left and right branches where $ P = P_l \ \Vert \ P_r$ and $P' = P_l' \ \Vert \ P_r'$.\\
			$(C_l, C_r, C_l', C_r')$ &  A set of ciphertext pairs with left and right branches where $C = C_l \ \Vert \ C_r$ and $C' = C_l' \ \Vert \ C_r'$.\\
			\bottomrule
	\end{tabular}}
\end{table}

\subsection{A Brief Description of \SIMON and \SIMECK Ciphers}
\textbf{\SIMON.} The lightweight family of AND-RX block ciphers \SIMON was proposed by the National Security Agency (NSA) in 2013. It adopts the Feistel structure and the round function consists of bitwise AND ($\odot)$, bitwise XOR ($\oplus$) and cyclic left shift $\gamma$ bit ($S^\gamma$) operation composition. The designer provides ten versions, all marked as \SIMONnospace$2n/mn$, where $2n$ represents the block size, $mn$ represents the key length, $n \in \{16,24,32,48,64\}$, $m \in \{2,3,4\}$. The round function of \SIMON algorithm is defined as:
\begin{equation*}
	f_{8,1,2}(x) = \left(S^8\left(x\right) \odot S^1\left(x\right)\right) \oplus S^2(x)\, .
\end{equation*}

The round keys are generated using a linear key schedule through the $K = (k_{m-1},k_{m-2},\ldots,k_0)$. A more complete description can refer to paper~\cite{beaulieu2015simon}.

\noindent\textbf{\SIMECK}. The \SIMECK family of lightweight block ciphers was designed by Yang \emph{et al}.~\cite{yang2015simeck}, aiming at improving the hardware implementation cost of \simon. \SIMECKnospace $2n/4n$ denotes an instance with a $2n$-bit block and a $4n$-bit key for $n\in \{16,24,32\}$. The round function of \SIMECK algorithm is defined as:
\begin{equation*}
	f_{5,0,1}(x) = \left(S^5\left(x\right) \odot S^0\left(x\right)\right) \oplus S^1(x)\, .
\end{equation*}

Conversely, \SIMECK uses the non-linear key schedule which reuses the cipher's round function to generate the round keys. A more complete description can be found in~\cite{yang2015simeck}.

\noindent\textbf{\SIMON-like ciphers.} Iterated ciphers that use \Simon's round function and generalize it to accept arbitrary rotational parameters are known as \Simon-like ciphers $(a,b,c)$. The \Simon-like function is then $f_{a,b,c}(x) = \left(S^a\left(x\right) \odot S^b\left(x\right)\right) \oplus S^c(x)$, which the rotational parameters $(a,b,c)$ are (8,1,2) and (5,0,1) for all \Simon and \Simeck versions, respectively.

\subsection{(Related-key) Differential Cryptanalysis}
Differential cryptanalysis is a chosen-plaintext attack introduced by Biham and Shamir in~\cite{biham1991differential}. It analyzes the effect of the difference of a plaintext pair on the difference of succeeding round outputs in an iterated cipher. Differential cryptanalysis is a widely used tool for the cryptanalysis of encryption algorithms and the development of new attacks due to its generality. Resistance to differential cryptanalysis became one of the basic criteria in the evaluation of the security of block ciphers.

\begin{definition}[Difference]\rm{~\cite{biham1991differential}} Let $X$ and $X'$ be two bit strings of length $n$, then the difference between $X$ and $X'$ is defined as: $\Delta X = X \oplus X'$.
\end{definition}

\begin{definition}[Differential Pair]\rm{~\cite{biham1991differential}}
	Let $\alpha, \beta$ be $n$-bit vectors, the difference value of the input pair $(X,X')$ of the block cipher is $X \oplus X' = \alpha$, after $r$-round of encryption, the difference value of the output pair $(Y,Y')$ is $Y \oplus Y' = \beta$, and let a round function $f: \mathbb{F}_2^n \to \mathbb{F}_2^n$, then $(\alpha, \beta)$ is called an $r$-round differential pair of block cipher, where $\alpha$ is the input difference of round function $f$, $\beta$ is the output difference of $f$. In particular, when $r = 1$, $(\alpha, \beta)$ characterizes the differential propagation characteristics of the round function $f$.
\end{definition}

For a specific cipher, the differential must be carefully selected to make the differential attack successful. This makes researchers need to study the internal process of the algorithm. The basic method is to track a path passed by a high probability differential at different stages of encryption. This is called differential characteristics in cryptography and is defined as follows.

\begin{definition}[Differential Characteristics]\rm{~\cite{biham1991differential}} Let $X, X'$ be $n$-bit vectors and $\beta _i$ be an $n$-bit constant. When the difference value of the input pair $(X, X')$ satisfies $X \oplus X' = \beta _0$, the difference value of the intermediate state $(Y_i,Y_i')$ satisfies $Y_i \oplus Y_i' = \beta _i$ during the $r$-th round of encryption, where, $1 \leq i \leq r$. Then, $\Omega = (\beta _1, \beta _2, \ldots, \beta _r)$ can be named an $r$-round differential characteristic of an iterative block cipher.
\end{definition}

For given differential characteristics, use the following definition to calculate its probability.
\begin{definition}\rm{~\cite{biham1991differential}}
	The probability $DP(\Omega)$ corresponding to an $r$-round differential characteristic $\Omega = (\beta _1, \beta _2, \ldots, \beta _r)$ of the iterative block cipher refers to the case where the input $X$ and the round keys are independent and random distributed, when the differential value of the input pair $(X,X')$ is $X \oplus X' = \beta_1$, in the $i$-round encryption process, the difference value of the intermediate state $(Y_i,Y_i')$ satisfies the probability of $Y_i \oplus Y_i' = \beta_i$, where $1 \leq i \leq r$. Under the above assumption, the probability of the differential characteristic is equal to the product of the differential propagation probabilities of each round, \emph{i.e.},:
	\begin{equation*}
		\begin{aligned}
			DP(\Omega) &= \prod \limits_{i=1}^r Pr(\beta _{i-1} \to \beta _i) \\
			&=  \prod \limits_{i=1}^r \frac{\{Y_{i-1} | f(Y_{i-1}) \oplus f(Y_{i-1} \oplus \beta _{i-1}) = \beta _i \}}{2^n}.
		\end{aligned}	
	\end{equation*}
\end{definition}

When the input difference undergoes a linear operation, it will be propagated through the operation with probability 1, and the output difference is deterministic, such as XOR ($\oplus$) and cyclic shift ($\lll, \ggg$) in the ARX operation. When the input difference passes through a non-linear operation, the difference propagation is often probabilistic. 

Related-key differential cryptanalysis was introduced by Biham in ~\cite{biham1994new}. Unlike the single-key differentials that have differences only in the plaintexts, related-key differential distinguishers have differences in the master keys as well. It exploits the output differences given a pair of plaintexts $P$ and $P'$ encrypted by a pair of related keys $K$ and $K'$, respectively. Related-keys differential cryptanalysis is also one of the basic criteria in the evaluation of the security of block ciphers, which has successfully attacked many block ciphers, such as~\cite{jakimoski2003related,ko2004related,biryukov2010automatic}.

\subsection{Convolutional Neural Network}
\textbf{Convolutional neural network (CNN)} is an important paradigm in deep learning. CNN is usually composed of the convolutional layer, non-linear layer, pooling layer and fully connected layer. According to the convolution dimension of the feature map, it can be divided into one-, two-, and three-dimensional convolutional neural network (\emph{i.e.}, 1D-CNN, 2D-CNN and 3D-CNN), where the 1D-CNN applies a convolution over a fixed (multi-)temporal input signal.

\textbf{Convolution Layer} (CONV). Convolution is the basic operation of CNN, and its main purpose is to extract features. The core task of CNN is to learn parameters to extract effective patterns. In the forward propagation, the training data will go through the convolution kernel with initial parameters to obtain the initial output. In the back propagation, a loss function will be applied to adjust the parameters to minimize the gap between the initial output and the target label. After several iterations, when the loss stabilizes, the training process will be finished. Note that in this paper we apply 1D-CNN, then the convolution layer can be denoted by Conv1D.

\textbf{Non-linear layer}. The main purpose of the non-linear layer is to introduce non-linear characteristics into the system. The most common non-linear layer in a CNN network is the rectified linear unit (\textbf{ReLU}) function, defined as $f(x) = max(0,x)$. Effectively, it removes negative values from an activation map by setting them to zero. It increases the nonlinear properties of the decision function and of the overall network without affecting the receptive fields of the convolution layer. Other functions are also used to increase nonlinearity, such as the sigmoid function. ReLU is often preferred to other functions because it trains the neural network several times faster without a significant penalty to generalization accuracy.

\textbf{Fully connected layer} (FC). The fully connected layer is generally located in the back layers of the network for performing the classification task. Usually, the input of the fully connected layer is the flatten feature map generated by convolution layer.

In addition, some functional layers may be used in CNN. For example, \textbf{Batch Normalization (BN)} can be applied after the convolution layer to reduce the internal covariate shift, which can effectively prevent the gradient disappearance problem and speed up network training.

\textbf{Residual Network (ResNet)} is one of the most representative CNNs, which was proposed by He \emph{et al.}~\cite{he2016deep} in 2015. ResNet can train a deeper CNN model to achieve higher accuracy. The core idea is to establish ``shortcuts (skip) connections'' between the front layer and the back layer. It is composed of a series of residual blocks. A residual block can be expressed as:
\[ x_{l+1} = x_l + \mathcal{F}(x_l).\]
It is divided into two parts: the direct mapping part and the residual part. $\mathcal{F}(x_l)$ is the residual part, which is generally composed of two or three convolution operations. The activation functions of ReLU and BN can be rearranged to create a variety of residual block variants.

\textbf{Squeeze-and-Excitation Network (SENet)} is a new network structure proposed by Hu \emph{et al.} that won the first place in ILSVRC 2017 classification competition~\cite{hu2018squeeze}. The ``Squeeze-and-Excitation'' (SE) block adaptively recalibrates channel-wise feature responses by explicitly modelling interdependencies between channels. It can be integrated into standard architectures by insertion after the non-linearity following each convolution. In this paper, SE block is used directly with the residual network, \emph{i.e.}, the SE-ResNet network.

\section{Improved (Related-key) Differential-based Neural Distinguishers}\label{sec:improve}

\subsection{Dateset: Multiple Ciphertext Pairs with New Data Format}
Data plays a very important role in deep learning, data preparation is a fundamental step for deep learning model development. Some researchers explored the use of multiple ciphertext pairs to improve the performance of differential-based neural distinguishers~\cite{benamira2021deeper, chen2021new, hou2021improve}. Some researchers also performed additional transformations on each pair of ciphertexts before feeding them into the network. Concretely, in Gohr's work, the $n$-round NDs fed with data of form $(C_l,C_r,C_l',C_r')$. Subsequently, Benamira \emph{et al.}~\cite{benamira2021deeper} conjected the first convolution layer of Gohr's neural network transforms the input $(C_l ,C_r ,C_l' ,C_r')$ into $(C_l \oplus C_l', C_l \oplus C_l' \oplus C_r \oplus C_r', C_l \oplus C_r, C_l' \oplus C_r')$ and a linear combination of those terms. In~\cite{hou2021improve}, Hou \emph{et al.} designed the NDs model with multiple output differences as a sample, \emph{i.e.}, the $n$-round NDs fed multiple pairs with data of form $(C_l \oplus C_l', C_r \oplus C_r') \triangleq (\Delta_L^{r}, \Delta_R^{r})$. In~\cite{bao2022enhancing}, Bao \emph{et al.} accepted the $r$-round NDs fed with data of form $(C_r, C_r', \Delta_R^{r-1})$, where $\Delta_R^{r-1} = ((C_r \lll 8) \odot (C_r \lll 1) \oplus (C_r \lll 2) \oplus C_l) \oplus ((C_r' \lll 8) \odot (C_r' \lll 1) \oplus (C_r' \lll 2) \oplus C_l')$ for \SIMON ciphers.

In this paper, we employ multiple ciphertext pairs with new data of form $(\Delta_L^{r}, \Delta_R^{r}, C_l, C_r, C_l' ,C_r', \Delta_R^{r-1}, p\Delta_R^{r-2})$ to improve the performance of neural distinguishers (the reason for choosing this data format is given in Section~\ref{subsec:exp-data}). Then, the process of constructing a dataset can be described.

For the differential-neural distinguisher, first encrypt the $s$ plaintext pairs $((P,P')^1,(P,P')^2,\ldots,(P,P')^s)$ with a random key to get the $s$ ciphertext pairs. Then, use the $s$ ciphertext pairs to get the data:
\begin{equation*}
	\begin{array}{*{20}{c}}
		{(\Delta_L^{r}, \Delta_R^{r}, C_l, C_r, C_l' ,C_r', \Delta_R^{r-1}, p\Delta_R^{r-2})^1,}\\
		{(\Delta_L^{r}, \Delta_R^{r}, C_l, C_r, C_l' ,C_r', \Delta_R^{r-1}, p\Delta_R^{r-2})^2,}\\
		\vdots \\
		{(\Delta_L^{r}, \Delta_R^{r}, C_l, C_r, C_l' ,C_r', \Delta_R^{r-1}, p\Delta_R^{r-2})^s}.
	\end{array}
\end{equation*}
where the set ${(\Delta_L^{r}, \Delta_R^{r}, C_l, C_r, C_l' ,C_r', \Delta_R^{r-1}, p\Delta_R^{r-2})^i}$ of row $i$ is denoted by $\Omega^i$.

Finally, splice $\Omega^i$ and convert it into a string of binary as a sample, and each sample will be attached a label $Y$:
\begin{equation*}
	\begin{aligned}
		\label{Ylable}
		 Y\left(\Omega^1 || \Omega^2 \cdots || \Omega^s \right)  =\left\{
		\begin{aligned}
			1 & ,  &\hbox{if} \ P^i \oplus (P')^i = \Delta_p, 1 \leq i \leq s,\\
			0 & ,  & \hbox{else}.
		\end{aligned}
		\right.
	\end{aligned}	
\end{equation*}
where $\Delta_p$ is a constant input difference. It examines how to select the $\Delta_p$ in Section~\ref{sec:input difference}.



Unlike differential-neural distinguisher, which uses a random key $K$ to encrypt the $s$ plaintext pairs, related-key differential-neural distinguisher uses a pair of keys $(K,K')$ with a difference of $\Delta_k$ to encrypt the $s$ plaintext pairs.

We construct the dataset based on the above steps and set $s = 8$. In the basic training process, the size of the training set is $2\times10^7$, and the test set is $2\times10^6$. Meanwhile, there is an independent key used for each sample. Therefore, the training set has $2\times10^7$ corresponding random keys, and the test set has $2\times10^6$ corresponding random keys.

\subsection{Network Architecture}
A deep learning architecture is a multilayer stack of simple modules, most of which are subject to learning, and many of which compute non-linear input-output mappings. Each module in the stack transforms its input to increase both the selectivity and the invariance of the representation. With multiple non-linear layers, say a depth of 5 to 20, a system can implement extremely intricate functions of its inputs that are simultaneously sensitive to minute details.

\begin{figure*}
	\centering
	\includegraphics[scale=0.7]{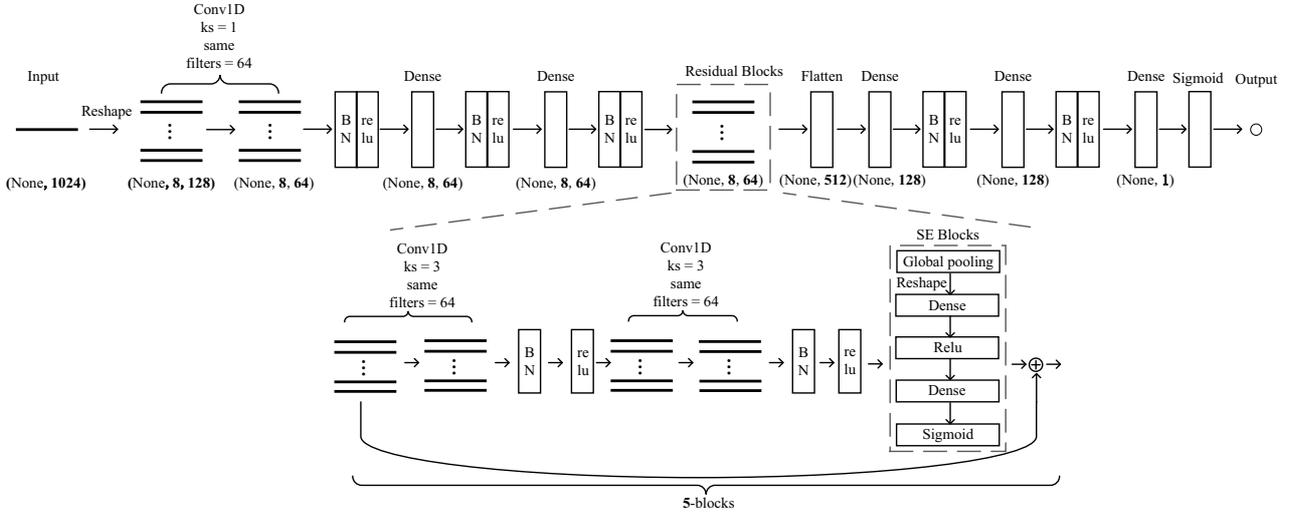}
	\caption{Network architecture proposed in this paper.}\label{fig:network}
\end{figure*}

Given the success of ResNet on \speck~\cite{gohr2019improving} and SENet on \simon~\cite{bao2022enhancing}, as well as their superior performance on classification tasks, we use the SE-ResNet network. As shown in Fig.~\ref{fig:network}, the network consists of three main components: input layer, iteration layer and predict layer. The input layer uses one Conv1D layer and two Dense layers to receive fixed length training data. In the iteration layer, use 5 SE-ResNet modules where each module contains two Conv1D layers and one SE block. To make the network learning more stable and alleviate the problem of gradient disappearance, a BN layer is applied after each Conv1D layer, and then followed by an activation layer with ReLU function. Finally, in predict layer, to make the data smoothly transform from the convolutional layer to the fully connected layer, we introduce a flatten layer to perform one-dimensional flattening of the data output from the convolutional layer. The fully connected layer consists of two Dense layers where each has 64 neurons and an output unit with only one neuron.

We set the batch size to 30000, cyclic learning rate $l_i=\alpha + \frac{(n-i)mod(n+1)}{n} \cdot (\beta -\alpha)$ with $\alpha=0.0001$, $\beta =0.003$, $n=29$ for epoch $i$, which is denoted as cyclic\dlmu[0.2cm]{}\!lr$(30,0.003,0.0001)$. Adam~\cite{kingma2014adam} is used as the optimizer with mean squared error (MSE) loss function and L2 regularization parameterized by $c=0.00001$. Each dataset is trained with 120 epochs for the basic training method. The accuracy, TPR, and TNR of the ND are the average results after 5 repetitions.

\section{Comparing the Performance of the Hybrid Distinguisher with Different Input Difference}\label{sec:input difference}
In this section, we investigate the effect of input difference on the performance of the hybrid distinguishers. Essentially, to be used in key-recovery, the NDs are too short such that they have to be prepended with classical differentials. Whether the resulting HDs can be used in a key-recovery attack depends on whether the input difference of NDs leads to better accuracy and, at the same time, leads to prepended CDs with high differential probability.

Therefore, taking the performance of the hybrid distinguisher of \SIMONnospace32/64 as an entry point, we investigate the issue in two phases. In the first stage, we study the performance of all input differences with Hamming weights of 1, 2, and 3 on the 11-round ND, and filter the input differences that can obtain a non-marginal advantage (accuracy above 0.50). Then study the performance of these filtered input differences on 12-round ND. In the second stage, we study the probability of the prepended CDs with these filtered input differences. 


\vspace{1em}

\noindent\textbf{The First Stage}
\vspace{1em}

\begin{figure*}
	\centering
	\includegraphics[scale=0.45]{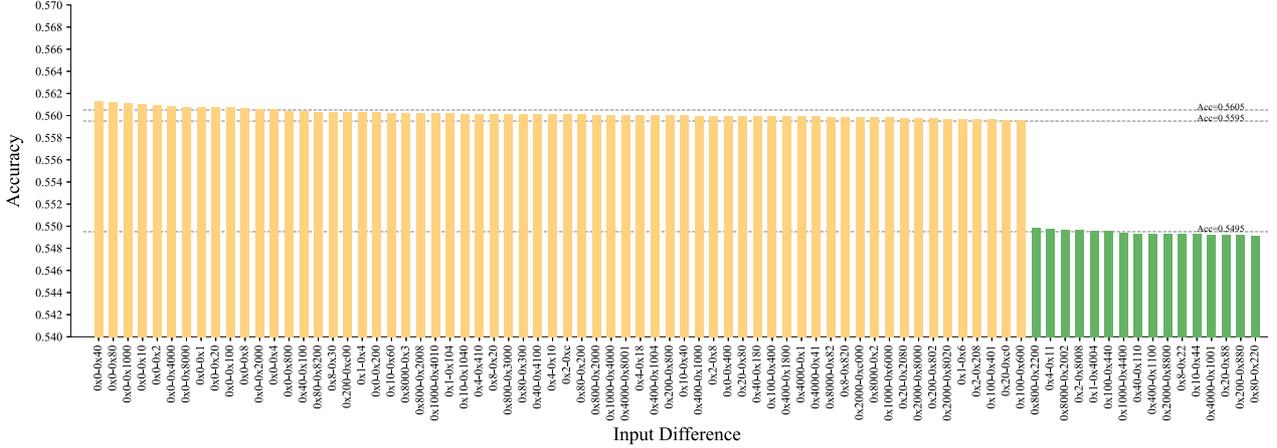}
	\caption{The input differences with Hamming weights of 1, 2, and 3 that can obtain a clear non-marginal advantage (accuracy above 0.52) on the 11-round ND of \SIMONnospace32/64 with 8 ciphertext pairs as a sample.}\label{fig:input-difference}
\end{figure*}

Let HW$(\Delta_p)$ denote the Hamming weight of the input difference, then there are $32+496+4960=5488$ input difference with HW$(\Delta_p) \leq 3$. Based on Section~\ref{sec:improve}, traversing these input difference $\Delta_p$ with the batch size 30000 and cyclic\dlmu[0.2cm]{}\!lr$(30,0.003,0.0001)$, we construct 11-round ND of \SIMONnospace32/64, respectively. There are 128 input differences filtered, of which 48 have an accuracy between 0.51-0.52 and 80 have an accuracy between 0.54-0.56. Therefore, we mainly focus on the performance of these 80 input differences. The results with these 80 input differences are shown in Fig.~\ref{fig:input-difference}.

It is discovered that 11-round ND with input difference $\Delta_p=(a,b)$ and input difference $\Delta_p'=(a\lll i,b\lll i)$ have similar accuracy, for $0 \leq i < 16$. Thus, we only list one of these 16 input differences in Table~\ref{tab:input difference perform}. Specifically, for HW$(\Delta_p)=1$, using the input difference (omit the $0x$ symbol):

\begin{center}
(0000,0001), (0000,0002), (0000,0004), (0000,0008),\\
(0000,0010), (0000,0020), (0000,0040), (0000,0080),\\
(0000,0100), (0000,0200), (0000,0400), (0000,8000),\\
(0000,1000), (0000,2000), (0000,4000), (0000,8000),
\end{center}
can construct 11-round ND of \SIMONnospace32/64 with an accuracy of about 0.561.

For HW$(\Delta_p)=2$, using the input difference:
\begin{center}
(0001,0004), (0002,0008), (0004,0010), (0008,0020),\\
(0010,0040), (0020,0080), (0040,0100), (0080,0200),\\
(0100,0400), (0200,0800), (0400,1000), (0800,2000),\\
(1000,4000), (2000,8000), (4000,0001), (8000,0002),
\end{center}
can build 11-round ND of \SIMONnospace32/64 with an accuracy of about 0.560.

For HW$(\Delta_p)=3$, there are three sets of $(a,b)$. Using the input difference:
\begin{center}
	(0001,0104), (0002,0208), (0004,0410), (0008,0820),\\
	(0010,1040), (0020,2080), (0040,4100), (0080,8200),\\
	(0100,0401), (0200,0802), (0400,1004), (0800,2008),\\
	(1000,4010), (2000,8020), (4000,0041), (8000,0082),
\end{center}
can construct 11-round ND of \SIMONnospace32/64 with an accuracy of about 0.560.

Using the input difference:
\begin{center}
	(0001,0006), (0002,000c), (0004,0018), (0008,0030),\\
	(0010,0060), (0020,00c0), (0040,0180), (0080,0300),\\
	(0100,0600), (0200,0c00), (0400,1800), (0800,3000),\\
	(1000,6000), (2000,c000), (4000,8001), (8000,0003),
\end{center}
can obtain 11-round ND of \SIMONnospace32/64 with an accuracy of about 0.560.

Using the input difference:
\begin{center}
	(0001,4004), (0002,8008), (0004,0011), (0008,0022),\\
	(0010,0044), (0020,0088), (0040,0110), (0080,0220),\\
	(0100,0440), (0200,0880), (0400,1100), (0800,2200),\\
	(1000,4400), (2000,8800), (4000,1001), (8000,2002),
\end{center}
can get 11-round ND of \SIMONnospace32/64 with an accuracy of about 0.549.

It can be found that the effect of the 16 input differeces \textbf{\texttt{(0x0001 $\lll i$,0x4004 $\lll i$)}} ($0 \leq i <16$) is slightly inferior to the other 64 ($80-16=64$) input differeces for 11-round ND.

Then, with the input differences \textbf{\texttt{(0x0,0x1)}, \texttt{(0x1,0x4)},\texttt{(0x1,0x104)},\texttt{(0x1,0x6)},\texttt{(0x1,0x4004)}} separately, we construct 12-round ND of \SIMONnospace32/64 by using the basic training method. The results are shown in Table~\ref{tab:12round}. It shows the accuracy exceeds 0.50 except for the input difference \textbf{\texttt{(0x0,0x1)}} (\textbf{\texttt{(0x0,0x1)}} can get an accuracy of 0.5142 by using the staged training method). Therefore, a total of 64 input differences can make 12-round ND obtain non-marginal advantage by using the basic training method. Meanwhile, the input differential \textbf{\texttt{(0x1,0x4)}} and \textbf{\texttt{(0x1,0x6)}} performed the best, with an accuracy of 0.5152.

\begin{table}
	\centering
	\renewcommand\arraystretch{1.1}
	\scalebox{1}{
		\newcommand{\tabincell}[2]{\begin{tabular}{@{}#1@{}}#2\end{tabular}}
		\begin{threeparttable}
			\caption{Experiment with Different Input Difference of 12-round ND for \SIMONnospace32/64 with 8 ciphertext pairs as a sample.}
			\begin{tabular}{p{0.9cm}<{\centering}p{2.6cm}<{\centering}p{0.9cm}<{\centering}p{0.9cm}<{\centering}p{0.9cm}<{\centering}}
				\toprule
				Cipher  &  Input Difference & Acc & TPR & TNR \\
				\midrule
				\multirow{5}{*}{\tabincell{c}{\SIMONnospace\\32/64}}   & \textbf{\texttt{(0x0000,0x0001)}}  & 0.5004  &0.1149 &0.8857 \\
				
				& \textbf{\texttt{(0x0001,0x0004)}} & 0.5152  & 0.4799  & 0.5505 \\	
				
				& \textbf{\texttt{(0x0001,0x0104)}} & 0.5151  & 0.4901  & 0.5401 \\	
				
				& \textbf{\texttt{(0x0001,0x0006)}} & 0.5152  & 0.4852   & 0.5453 \\	
				
				& \textbf{\texttt{(0x0001,0x4004)}} & 0.5135  & 0.4331  & 0.5940 \\			
				\bottomrule
			\end{tabular}
			\label{tab:12round}
	\end{threeparttable}}	
\end{table}

\vspace{1em}

\noindent\textbf{The Second Stage}
\vspace{1em}

The NDs are prepended with 3 rounds of CDs in~\cite{bao2022enhancing}, so we use 3 rounds prepended CDs as a benchmark to test the performance of the input differential filtered in the first stage. An SMT solver is used to determine the probability of prepended CDs. We first decide if a differential characteristic with probability $p$ exists, then enumerate all differential characteristics with a probability of $p$. The results are presented in Table~\ref{tab:input difference perform}. It can be seen that the probability of the 3 rounds prepended CDs with the input difference \textbf{\texttt{(0x0000 $\lll i$,0x0001 $\lll i$)}}, $0 \leq i <16$ (\emph{i.e.}, \textbf{\texttt{(0,$e_i$)}}) are the highest, followed by 2-bit input differential \textbf{\texttt{(0x0001 $\lll i$,0x0004 $\lll i$)}}, $0 \leq i <16$, and the worst are \textbf{\texttt{(0x0001 $\lll i$,0x4004 $\lll i$)}}, $0 \leq i <16$.

As a result, after these two steps of filtering, the input difference \textbf{\texttt{(0,$e_i$)}} is possibly the best option for hybrid distinguishers. Meanwhile, the input difference \textbf{\texttt{(0x0001 $\lll i$,0x0004 $\lll i$)}}, $0 \leq i <16$  is also a good choice. But we cannot yet give a clearer opinion on how much $i$ is set.

\begin{table*}[htbp]
	\centering
	\renewcommand\arraystretch{1}
	\newcommand{\tabincell}[2]{\begin{tabular}{@{}#1@{}}#2\end{tabular}}
	\caption{Comparing the performance of the hybrid distinguisher with different input difference for \SIMONnospace32/64. The NDs is 11-round. The number on the arrow represents the probability of the differential characteristic from the input difference to the output difference, and the number of characteristics. For example: $\left. {\begin{array}{*{20}{r}}
		{\textbf{\texttt{(0011,0040)}}} \\ 
		{...} 
\end{array}} \right\}\xrightarrow{{{2^{ - 8}}\left( {\# 20} \right)}}{\textbf{\texttt{(0000,0001)}}}$ means that when the input differences are \textbf{\texttt{(0000,0001)}} etc. and the output difference is \textbf{\texttt{(0000,0001)}}, there are 20 characteristics with a probability of $2^{-8}$ for 3-round \SIMONnospace32/64. And these input differences in the prepended CDs are the smallest Hamming weight in these characteristics.}
	\begin{threeparttable}
		
		\begin{tabular}{p{1.5cm}<{\centering}p{2cm}<{\centering}p{1cm}<{\centering}p{1cm}<{\centering}p{1cm}<{\centering}p{7cm}<{\centering}}
			\toprule
			HW$(\Delta_p)$ & $\Delta_p$ & ND's Acc & ND's TPR & ND's TNR & Prepended CDs (3-round) \\
			\midrule
			1-bit 
			& \textbf{\texttt{(0000,0001)}} & 0.5607 & 0.5407 & 0.5807 & $\left. {\begin{array}{*{20}{r}}
					{\left. {\begin{array}{*{20}{r}}
								{\textbf{\texttt{(0011,0040)}}}\\
								{...}
						\end{array}} \right\}\xrightarrow[]{2^{-8}(\#20)}}\\
					{\left. {\begin{array}{*{20}{r}}
								{\textbf{\texttt{(0010,0146)}}}\\
								{...}
						\end{array}} \right\}\xrightarrow[]{2^{-9}(\#4)}}\\
					{\left. {\begin{array}{*{20}{r}}
								{\textbf{\texttt{(0011,0040)}}}\\
								{...}
						\end{array}} \right\}\xrightarrow[]{2^{-10}(\#232)}}\\
					{\left. {\begin{array}{*{20}{r}}
								{\textbf{\texttt{(0011,0040)}}}\\
								{...}
						\end{array}} \right\}\xrightarrow[]{2^{-11}(\#352)}}\\
					
					...
			\end{array}} \right\} \textbf{\texttt{(0000,0001)}}$ 
			
			 \\			
			
			
			\midrule	
			
			2-bit 
			& \textbf{\texttt{(0001,0004)}} & 0.5602 & 0.5059 & 0.6145 & $\left. {\begin{array}{*{20}{r}}
					{\left. {\begin{array}{*{20}{r}}
								{\textbf{\texttt{(0040,0111)}}}\\
								{...}
						\end{array}} \right\}\xrightarrow[]{2^{-8}(\#4)}}\\
					{\left. {\begin{array}{*{20}{r}}
								{\textbf{\texttt{(0140,0511)}}}\\
								{...}
						\end{array}} \right\}\xrightarrow[]{2^{-9}(\#10)}}\\
					{\left. {\begin{array}{*{20}{r}}
								{\textbf{\texttt{(1040,0101)}}}\\
								{...}
						\end{array}} \right\}\xrightarrow[]{2^{-10}(\#72)}}\\
					{\left. {\begin{array}{*{20}{r}}
								{\textbf{\texttt{(1060,0101)}}}\\
								{...}
						\end{array}} \right\}\xrightarrow[]{2^{-11}(\#124)}}\\
					
					...
			\end{array}} \right\} \textbf{\texttt{(0001,0004)}}$ \\

			\midrule
			
			
			\multirow{20}{*}{\tabincell{c}{3-bit}} 
			& \textbf{\texttt{(0001,0104)}} & 0.5601 & 0.5024 & 0.6179 & $\left. {\begin{array}{*{20}{r}}
					{\left. {\begin{array}{*{20}{r}}
								{\textbf{\texttt{(0040,0111)}}}\\
								{...}
						\end{array}} \right\}\xrightarrow[]{2^{-10}(\#4)}}\\
					{\left. {\begin{array}{*{20}{r}}
								{\textbf{\texttt{(0140,0110)}}}\\
								{...}
						\end{array}} \right\}\xrightarrow[]{2^{-11}(\#48)}}\\
					{\left. {\begin{array}{*{20}{r}}
								{\textbf{\texttt{(1040,0101)}}}\\
								{...}
						\end{array}} \right\}\xrightarrow[]{2^{-12}(\#80)}}\\
					{\left. {\begin{array}{*{20}{r}}
								{\textbf{\texttt{(0200,4201)}}}\\
								{...}
						\end{array}} \right\}\xrightarrow[]{2^{-13}(\#620)}}\\

					...
			\end{array}} \right\} \textbf{\texttt{(0001,0104)}}$ \\

			\cmidrule{2-6}
			
			
			& \textbf{\texttt{(0001,0006)}} & 0.5597 & 0.4972 & 0.6221 &$\left. {\begin{array}{*{20}{r}}
					{\left. {\begin{array}{*{20}{r}}
								{\textbf{\texttt{(0040,0111)}}}\\
								{...}
						\end{array}} \right\}\xrightarrow[]{2^{-10}(\#4)}}\\
					{\left. {\begin{array}{*{20}{r}}
								{\textbf{\texttt{(0140,0511)}}}\\
								{...}
						\end{array}} \right\}\xrightarrow[]{2^{-11}(\#8)}}\\
					{\left. {\begin{array}{*{20}{r}}
								{\textbf{\texttt{(1040,0101)}}}\\
								{...}
						\end{array}} \right\}\xrightarrow[]{2^{-12}(\#72)}}\\
					{\left. {\begin{array}{*{20}{r}}
								{\textbf{\texttt{(1060,0101)}}}\\
								{...}
						\end{array}} \right\}\xrightarrow[]{2^{-13}(\#296)}}\\
					
					...
			\end{array}} \right\} \textbf{\texttt{(0001,0006)}}$ \\

			\cmidrule{2-6}
			
			& \textbf{\texttt{(0001,4004)}} & 0.5495 & 0.4433 & 0.6557 &$\left. {\begin{array}{*{20}{r}}
					{\left. {\begin{array}{*{20}{r}}
								{\textbf{\texttt{(4044,0101)}}}\\
								{...}
						\end{array}} \right\}\xrightarrow[]{2^{-12}(\#80)}}\\
					{\left. {\begin{array}{*{20}{r}}
								{\textbf{\texttt{(4064,0101)}}}\\
								{...}
						\end{array}} \right\}\xrightarrow[]{2^{-13}(\#344)}}\\
					{\left. {\begin{array}{*{20}{r}}
								{\textbf{\texttt{(0144,0140)}}}\\
								{...}
						\end{array}} \right\}\xrightarrow[]{2^{-14}(\#1072)}}\\
					
					...
			\end{array}} \right\}  \textbf{\texttt{(0001,4004)}}$ \\

			\bottomrule
		\end{tabular}
	\end{threeparttable}
	\label{tab:input difference perform}	
\end{table*}

\section{(Related-key) Differential-Neural Distinguishers for Round-Reduced \SIMONnospace32/64 and \SIMONnospace64/128}\label{sec:app-simon}
In this section, the NDs are trained using the basic training method and the staged training method. The training model is based on Section~\ref{sec:improve}.

\subsection{Differential-Neural Distinguishers}
\textbf{\SIMONnospace32/64}
\vspace{1em}

\noindent\textbf{Training using the basic scheme.} Using the input difference $\textbf{\texttt{(0x0000,0x0040)}}$, we build NDs against \SIMONnospace32/64 cover to 9-, 10-, and 11-round with 0.9176, 0.6975, and 0.5609 accuracy, respectively. Using the input difference $\textbf{\texttt{(0x0001,0x0004)}}$, we build 12-round ND with 0.5152 accuracy. Table~\ref{tab:overview of results} presents the results.

Note that for NDs fed with single ciphertext pairs, with multiple ciphertext pairs with the same label, one can directly obtain a combine-response distinguisher (CRD) using the formula (3) in~\cite{gohr2019improving}. Similar to the NDs fed with multiple ciphertext pairs, the CRDs' accuracy improves quickly with increasing the number of ciphertext pairs. Therefore, we compare the accuracy of NDs with CRDs under the number of ciphertext pairs with the same label. Compared with~\cite{bao2022enhancing}, the accuracy of our NDs are improved.

\vspace{1em}

\noindent\textbf{Training using the Staged Training Method.} We also use several stages of pre-training to train a 12-round differential-neural distinguisher for \SIMONnospace32/64. In the first stage, the best 10-round distinguisher is retained to recognize 9-round \SIMONnospace32/64 with the input difference  $\textbf{\texttt{(0x0440,0x0100)}}$. The number of samples for training and for testing are $2^{25}$ and $2^{23}$, respectively. The number of epochs is 30 and the learning rate is $10^{-4}$. 

In the second stage, the best network of the first stage is retained to recognize 12-round \SIMONnospace32/64 with the input difference $\textbf{\texttt{(0x0000,0x0040)}}$. For this stage, $2^{25}$ and $2^{23}$ examples are freshly generated for training and testing, respectively. The learning rate is $10^{-4}$ for 30 epochs.

Cyclical learning rates are also used for these training stages, the first and second stage both use a minimum learning rate of 0.0001 and a maximum of 0.001. All cycle lengths in these stages are set to 30 epochs. Eventually, the resulting ND achieves an accuracy of 0.5142.

\vspace{1em}

\noindent\textbf{\SIMONnospace64/128}
\vspace{1em}

\noindent\textbf{Training using the basic scheme.} Based on the input difference $\textbf{\texttt{(0x00000000,0x00000040)}}$, the NDs reach 0.9181, 0.7117, 0.5722, and 0.5148 accuracy for 11-, 12-, 13-, and 14-round, respectively. As shown in Table~\ref{tab:overview of results}, the results are summarized.

\vspace{1em}
\noindent\textbf{Training using the Staged Training Method.}  The best 14-round distinguisher for \SIMONnospace64/128 is trained using the staged training method.  

In the first stage, the retained best 12-round distinguisher is trained and tested with 11-round  $2^{25}$ and $2^{23}$ samples of \SIMONnospace64/128 with the input difference \textbf{\texttt{(0x00000440,0x00000100)}}. The number of epochs is 30 and the learning rate is $10^{-4}$.  The learning rate scheduler used in this stage is cyclic\dlmu[0.2cm]{}\!lr$(30,0.001,0.0001)$. 

Then the best network from the first stage is trained in the second stage. The number of examples for training and for testing are $2^{25}$ and $2^{23}$, using 14-round \SIMONnospace64/128 data with the input difference $\textbf{\texttt{(0x00000000,0x00000040)}}$. This stage is done in 30 epochs with learning rate of $10^{-4}$. The learning rate scheduler used in this stage is cyclic\dlmu[0.2cm]{}\!lr$(30,0.001,0.0001)$.
Finally, the accuracy of the resulting ND is 0.5185.

\vspace{1em}

\subsection{Related-key Differential-Neural Distinguishers}
We use the basic training method to train the related-key differential-neural distinguishers. Based on the plaintext difference $\textbf{\texttt{(0x0000,0x0040)}}$ and the key difference $\textbf{\texttt{(0x0000,0x0000,0x0000,0x0040)}}$, we enjoy 1, 0.9604, 0.6477, and 0.5262 accuracy for 10-, 11-, 12-, and 13-round RKNDs against \SIMONnospace32/64, respectively. 

Based on the plaintext difference $\textbf{\texttt{(0x00000000,0x00000040)}}$ and the key difference $ \textbf{\texttt{(0x00000000,0x00000000,0x00000000,0x00000040)}}$, we build RKNDs cover to 12-, 13-, and 14-round with 0.9880, 0.8398, and 0.5788 accuracy for \SIMONnospace64/128, respectively. To the best of our knowledge, this is the first successful application of the RKNDs against \Simon-like ciphers.

\subsection{Experiment with Different Data Format}\label{subsec:exp-data}
In order to improve the accuracy of the ND, we introduce a new data format $(\Delta_L^{r}, \Delta_R^{r}, C_l, C_r, C_l' ,C_r', \Delta_R^{r-1}, p\Delta_R^{r-2})$ suitable for the network architecture in this paper. Here, we explain the reason for choosing this data format. We mainly compare the effect of the different data format on the performance of the network based on the experiment of 9-, 10-, and 11-round NDs for \SIMONnospace32/64.

We use the basic method to train the 9-, 10-, and 11-round NDs based the input difference $\textbf{\texttt{(0x0000,0x0040)}}$, batch size 30000, and cyclic\dlmu[0.2cm]{}\!lr$(30,0.003,0.0001)$. The results are presented in Table~\ref{tab:data-format}.

It shows that the NDs using data formats of $(C_r, C_r', \Delta_R^{r-1})$, $(\Delta_L^{r}, \Delta_R^{r}, C_l, C_r, C_l' ,C_r', \Delta_R^{r-1})$, $(\Delta_L^{r}, \Delta_R^{r}, C_l, C_r, C_l' ,C_r', \Delta_R^{r-1}, p\Delta_R^{r-2})$ can achieve 11-round, and the accuracy with data format $(\Delta_L^{r}, \Delta_R^{r}, C_l, C_r, C_l' ,C_r', \Delta_R^{r-1}, p\Delta_R^{r-2})$ is greater than others. This is the primary cause for using this data format in the paper.

Meanwhile, it is noted that the accuracy dropped when the $p\Delta_R^{r-2}$ component was deleted from the data format $(\Delta_L^{r}, \Delta_R^{r}, C_l, C_r, C_l' ,C_r', \Delta_R^{r-1}, p\Delta_R^{r-2})$, \emph{i.e.}, the neural network benefits from providing data $p\Delta_R^{r-2}$. In fact, $p\Delta_R^{r-2}$ denotes the partial $\Delta_R^{r-2}$, and it can be determined without the round key when the ciphertext pair is given.

It is important to note that this comparison is only to show that the data format used in this paper better matches the current network for better performance. Different results may occur when the network is changed.

\begin{table*}
	\centering
	\renewcommand\arraystretch{1.1}
	\scalebox{1}{
		\newcommand{\tabincell}[2]{\begin{tabular}{@{}#1@{}}#2\end{tabular}}
		\begin{threeparttable}
			\caption{Experiment with different data format of 9-, 10-, and 11-round NDs for \SIMONnospace32/64. The best NDs for 9-, 10-, and 11-round are shown shaded.}
			\begin{tabular}{p{1cm}<{\centering}p{1cm}<{\centering}p{5.8cm}<{\centering}p{1.4cm}<{\centering}p{1.4cm}<{\centering}p{1.4cm}<{\centering}p{1.8cm}<{\centering}}
				\toprule
				Cipher  &  Round &  Data Format & Acc & TPR & TNR & Source\\
				\midrule
				\multirow{15}{*}{\tabincell{c}{\SIMONnospace\\32/64}}  & \multirow{5}{*}{9} & $(C_l,C_r,C_l',C_r')$  & 0.7524 & 0.7304 & 0.7743 &~\cite{gohr2019improving} \\
			    &  &  $(\Delta_L^{r}, \Delta_R^{r})$ & 0.6895 & 0.6613 & 0.7176 &~\cite{hou2021improve} \\
			    &  & $(C_r, C_r', \Delta_R^{r-1})$  & 0.8908 & 0.8786 & 0.9031 &~\cite{bao2022enhancing} \\	
			    &  & $(\Delta_L^{r}, \Delta_R^{r}, C_l, C_r, C_l' ,C_r', \Delta_R^{r-1})$ & 0.8945 & 0.8834 & 0.9057 & This Paper. \\
			    &  & $(\Delta_L^{r}, \Delta_R^{r}, C_l, C_r, C_l' ,C_r', \Delta_R^{r-1}, p\Delta_R^{r-2})$ \cellcolor{mygray}& 0.9176 \cellcolor{mygray}& 0.9052 \cellcolor{mygray}& 0.9299 \cellcolor{mygray}& \cellcolor{mygray}This Paper. \\

			    \cline{2-7}
			    
			    & \multirow{5}{*}{10} & $(C_l,C_r,C_l',C_r')$ & 0.5007 & 0.7015 & 0.2989 & ~\cite{gohr2019improving} \\	
			    
			    &  & $(\Delta_L^{r}, \Delta_R^{r})$ & 0.5605 & 0.5402 & 0.5809 & ~\cite{hou2021improve}\\	
			    
			    & & $(C_r, C_r', \Delta_R^{r-1})$ & 0.6856 & 0.6610 & 0.7102 & ~\cite{bao2022enhancing} \\	
			    & & $(\Delta_L^{r}, \Delta_R^{r}, C_l, C_r, C_l' ,C_r', \Delta_R^{r-1})$ & 0.6889 & 0.6639 & 0.7139 &This Paper.  \\	
			    & & $(\Delta_L^{r}, \Delta_R^{r}, C_l, C_r, C_l' ,C_r', \Delta_R^{r-1}, p\Delta_R^{r-2})$ \cellcolor{mygray}& 0.6975 \cellcolor{mygray}& 0.6662 \cellcolor{mygray}& 0.7287\cellcolor{mygray} & \cellcolor{mygray}This Paper. \\

			    \cline{2-7}
			    
			    & \multirow{5}{*}{11} & $(C_l,C_r,C_l',C_r')$ & 0.5006 & 0.4148 & 0.5863 & ~\cite{gohr2019improving} \\	
			    
			    &  & $(\Delta_L^{r}, \Delta_R^{r})$ & 0.5007 & 0.8110 & 0.1898 & ~\cite{hou2021improve}\\	
			    
			    & & $(C_r, C_r', \Delta_R^{r-1})$ & 0.5555 & 0.5437 & 0.5673 & ~\cite{bao2022enhancing} \\	
			    & & $(\Delta_L^{r}, \Delta_R^{r}, C_l, C_r, C_l' ,C_r', \Delta_R^{r-1})$ & 0.5578 & 0.5455 & 0.5700 &This Paper.  \\	
			    & & $(\Delta_L^{r}, \Delta_R^{r}, C_l, C_r, C_l' ,C_r', \Delta_R^{r-1}, p\Delta_R^{r-2})$ \cellcolor{mygray}& 0.5609 \cellcolor{mygray}& 0.5366\cellcolor{mygray} & 0.5852 \cellcolor{mygray}& \cellcolor{mygray}This Paper. \\
				\bottomrule
			\end{tabular}
	\label{tab:data-format}
	\end{threeparttable}}	
\end{table*}

\section{(Related-key) Differential-Neural Distinguishers for Round-Reduced \SIMECKnospace32/64 and \SIMECKnospace64/128}\label{sec:app-simeck}
\Simeck is a lightweight block cipher family that combines the good design components of \Simon and \Speck to make it even more compact and efficient. In this section, we build NDs and RKNDs for round-reduced \SIMECKnospace32/64 and \SIMECKnospace64/128.

\subsection{Differential-Neural Distinguishers}
\textbf{\SIMECKnospace32/64}
\vspace{1em}

\noindent\textbf{Training using the basic scheme.} Using the input difference $\textbf{\texttt{(0x0000,0x0040)}}$, we build NDs against \SIMECKnospace32/64 cover to 9-, 10-, and 11-round with 0.9952, 0.7354, and 0.5646 accuracy, respectively. The results are presented in Table~\ref{tab:overview of results}.

\vspace{1em}

\noindent\textbf{Training using the Staged Training Method.} A 12-round differential-neural distinguisher for \SIMECKnospace32/64 is also obtained by utilizing several stages of pre-training.

The first stage selects the best 10-round distinguisher to recognize 9-round \SIMECKnospace32/64 with the input difference $\textbf{\texttt{(0x0140,0x0080)}}$. Note that the most likely difference to appear three rounds after the input difference $\textbf{\texttt{(0x0000,0x0040)}}$ is $\textbf{\texttt{(0x0140,0x0080)}}$, and the probability is about $2^{-4}$.

It freshly generates $2^{25}$ and $2^{23}$ samples to train and test the distinguisher, respectively. This stage has 30 epochs and a learning rate of $10^{-4}$. The learning rate scheduler used in this stage is cyclic\dlmu[0.2cm]{}\!lr$(30,0.001,0.0001)$.

The best network obtained from the first stage is retained to recognize 12-round \SIMECKnospace32/64 with the input difference $\textbf{\texttt{(0x0000,0x0040)}}$. The number of examples for training and for testing are $2^{25}$ and $2^{23}$, respectively. The number of epochs is 30 and the learning rate is $10^{-4}$. The learning rate scheduler used in this stage is cyclic\dlmu[0.2cm]{}\!lr$(30,0.001,0.0001)$. Lastly, the ND produced has an accuracy of 0.5146.

\vspace{1em}

\noindent\textbf{\SIMECKnospace64/128}

\vspace{1em}

\noindent\textbf{Training using the basic scheme.} Similarly, based on the input difference $\textbf{\texttt{(0x00000000,0x00000040)}}$, the NDs reach accuracies of 0.9142, 0.7663, 0.6356, 0.5577, and 0.5202 for 14-, 15-, 16-, 17-, and 18-round, respectively. The results are shown in Table~\ref{tab:overview of results}.

\vspace{1em}
\noindent\textbf{Training using the Staged Training Method.} We use the staged training method to obtain the best 18-round distinguisher for \SIMECKnospace64/128.  

In the first stage, the retained best 16-round distinguisher is trained and tested with 15-round  $2^{25}$ and $2^{23}$ samples of \SIMECKnospace64/128 with the input difference \textbf{\texttt{(0x0000140,0x00000080)}}. The number of epochs is 30 and the learning rate is $10^{-4}$.  

Then the best network from the first stage is trained in the second stage. The number of freshly generated examples for training and for testing are $2^{25}$ and $2^{23}$, using 18-round \SIMECKnospace64/128 data with the input difference $\textbf{\texttt{(0x00000000,0x00000040)}}$. This stage is done in 30 epochs with learning rate of $10^{-4}$. 

Cyclical learning rates are used for these training stages, the first and second stage both use a minimum learning rate of 0.0001 and a maximum of 0.001. All cycle lengths in these stages are set to 30 epochs. As a final result, the ND produced has an accuracy of 0.5218.

\vspace{1em}

\subsection{Related-key Differential-Neural Distinguishers}
For related-key differential-neural distinguishers, based on the input difference $\textbf{\texttt{(0x0000,0x0040)}}$ and the key difference $\textbf{\texttt{(0x0000,0x0000,0x0000,0x0040)}}$, it covers to 13-, 14-, and 15-round with 0.9950, 0.6679 and 0.5467 accuracy for \SIMECKnospace32/64, respectively.

For \SIMECKnospace64/128, based on the input difference $\textbf{\texttt{(0x00000000,0x00000040)}}$ and the key difference $\textbf{\texttt{(0x00000000,0x00000000,0x00000000,0x00000040)}}$, it cover to 18-, 19-, 20-, 21-, and 22-round with 0.9066, 0.7558, 0.6229, 0.5519, and 0.5180 accuracy for \SIMECKnospace64/128, respectively. It can be seen the gap of RKNDs for \SIMON and \SIMECK is obvious, and \Simon's key-expansion algorithm offers better resistance. This is consistent with the conclusion that Lu \emph{et al.} get using rotational-XOR cryptanalysis in~\cite{lu2021effect}.

\section{Conclusion}\label{sec:conclusion}
In this paper, we provide an in-depth analysis of the (related-key) differential-neural distinguishers for \SIMON and \SIMECK ciphers. We adopt the multiple ciphertext pairs with data of the form $(\Delta_L^{r}, \Delta_R^{r}, C_l, C_r, C_l' ,C_r', \Delta_R^{r-1}, p\Delta_R^{r-2})$ fed to the neural network to improve the accuracy of the neural distinguisher. Meanwhile, we investigate the impact of input difference on the performance of the hybrid distinguishers to select the appropriate input difference. For \SIMONnospace32/64, \SIMONnospace64/128, \SIMECKnospace32/64 and \SIMECKnospace64/128, we construct the (related-key) differential-neural distinguishers with higher accuracy.

It is undeniable that there are many factors that can affect the performance of neural distinguishers. This paper explores its impact on the performance of neural distinguishers from the perspective of data format and input difference. In the future, we plan to further explore ways that can improve the performance of neural networks from multiple dimensions, such as using methods of feature engineering to extract more essential features of the training data and so on.


\ack{This work was supported in part by the National Key Research and Development Program of China [No.2021YFB3100800]; and the State Key Laboratory of Information Security [2020-MS-02]; and the National Natural Science Foundation of China [grant numbers 61872379, 61702537]; and the Academy of Finland [grant number 331883].}

\section*{Data availability}
The data underlying this article are available in the article and in its online supplementary material.


\bibliographystyle{compj}
\bibliography{ref}

\begin{thebibliography}{99}

\bibitem{biham1991differential}
Biham, E. and Shamir, A. Differential cryptanalysis of des-like cryptosystems.
\newblock {\em Journal of CRYPTOLOGY}, {\bf  4}, 3--72.

\bibitem{matsui1993linear}
Matsui, M. Linear cryptanalysis method for des cipher.
\newblock {\em Workshop on the Theory and Application of of Cryptographic
  Techniques},  pp. 386--397. Springer.

\bibitem{knudsen2002integral}
Knudsen, L. and Wagner, D. Integral cryptanalysis.
\newblock {\em International Workshop on Fast Software Encryption},  pp.
  112--127. Springer.

\bibitem{bogdanov2014linear}
Bogdanov, A. and Rijmen, V. Linear hulls with correlation zero and linear
  cryptanalysis of block ciphers.
\newblock {\em Designs, codes and cryptography}, {\bf  70}, 369--383.

\bibitem{mouha2011differential}
Mouha, N., Wang, Q., Gu, D., and Preneel, B. Differential and linear
  cryptanalysis using mixed-integer linear programming.
\newblock {\em International Conference on Information Security and
  Cryptology},  pp. 57--76. Springer.

\bibitem{sun2014automatic}
Sun, S., Hu, L., Wang, P., Qiao, K., Ma, X., and Song, L. Automatic security
  evaluation and (related-key) differential characteristic search: application
  to simon, present, lblock, des (l) and other bit-oriented block ciphers.
\newblock {\em International Conference on the Theory and Application of
  Cryptology and Information Security},  pp. 158--178. Springer.

\bibitem{mouha2013proof}
Mouha, N. and Preneel, B. A proof that the arx cipher salsa20 is secure against
  differential cryptanalysis.
\newblock {\em IACR Cryptol. ePrint Arch.}, {\bf  2013}, 328.

\bibitem{kolbl2015observations}
K{\"o}lbl, S., Leander, G., and Tiessen, T. Observations on the simon block
  cipher family.
\newblock {\em Annual Cryptology Conference},  pp. 161--185. Springer.

\bibitem{minier2014solving}
Minier, M., Solnon, C., and Reboul, J. Solving a symmetric key cryptographic
  problem with constraint programming.
\newblock {\em ModRef 2014, Workshop of the CP 2014 Conference} ~13.

\bibitem{gerault2016constraint}
Gerault, D., Minier, M., and Solnon, C. Constraint programming models for
  chosen key differential cryptanalysis.
\newblock {\em International Conference on Principles and Practice of
  Constraint Programming},  pp. 584--601. Springer.

\bibitem{lecun2015deep}
LeCun, Y., Bengio, Y., and Hinton, G. Deep learning.
\newblock {\em nature}, {\bf  521}, 436--444.

\bibitem{bengio2021deep}
Bengio, Y., Lecun, Y., and Hinton, G. Deep learning for ai.
\newblock {\em Communications of the ACM}, {\bf  64}, 58--65.

\bibitem{rivest1991cryptography}
Rivest, R.~L. Cryptography and machine learning.
\newblock {\em International Conference on the Theory and Application of
  Cryptology},  pp. 427--439. Springer.

\bibitem{maghrebi2016breaking}
Maghrebi, H., Portigliatti, T., and Prouff, E. Breaking cryptographic
  implementations using deep learning techniques.
\newblock {\em International Conference on Security, Privacy, and Applied
  Cryptography Engineering},  pp. 3--26. Springer.

\bibitem{hospodar2011machine}
Hospodar, G., Gierlichs, B., De~Mulder, E., Verbauwhede, I., and Vandewalle, J.
  Machine learning in side-channel analysis: a first study.
\newblock {\em Journal of Cryptographic Engineering}, {\bf  1}, 293.

\bibitem{gohr2019improving}
Gohr, A. Improving attacks on round-reduced speck32/64 using deep learning.
\newblock {\em Annual International Cryptology Conference},  pp. 150--179.
  Springer.

\bibitem{benamira2021deeper}
Benamira, A., Gerault, D., Peyrin, T., and Tan, Q.~Q. A deeper look at machine
  learning-based cryptanalysis.
\newblock {\em Annual International Conference on the Theory and Applications
  of Cryptographic Techniques},  pp. 805--835. Springer.

\bibitem{bao2022enhancing}
Bao, Z., Guo, J., Liu, M., Ma, L., and Tu, Y. Enhancing differential-neural
  cryptanalysis.
\newblock {\em International Conference on the Theory and Application of
  Cryptology and Information Security}. Springer.

\bibitem{beaulieu2015simon}
Beaulieu, R., Shors, D., Smith, J., Treatman-Clark, S., Weeks, B., and Wingers,
  L. The simon and speck lightweight block ciphers.
\newblock {\em Proceedings of the 52nd Annual Design Automation Conference},
  pp. 1--6.

\bibitem{yang2015simeck}
Yang, G., Zhu, B., Suder, V., Aagaard, M.~D., and Gong, G. The simeck family of
  lightweight block ciphers.
\newblock {\em International Workshop on Cryptographic Hardware and Embedded
  Systems},  pp. 307--329. Springer.

\bibitem{biham1994new}
Biham, E. New types of cryptanalytic attacks using related keys.
\newblock {\em Journal of Cryptology}, {\bf  7}, 229--246.

\bibitem{jakimoski2003related}
Jakimoski, G. and Desmedt, Y. Related-key differential cryptanalysis of 192-bit
  key aes variants.
\newblock {\em International Workshop on Selected Areas in Cryptography},  pp.
  208--221. Springer.

\bibitem{ko2004related}
Ko, Y., Hong, S., Lee, W., Lee, S., and Kang, J.-S. Related key differential
  attacks on 27 rounds of xtea and full-round gost.
\newblock {\em International Workshop on Fast Software Encryption},  pp.
  299--316. Springer.

\bibitem{biryukov2010automatic}
Biryukov, A. and Nikoli{\'c}, I. Automatic search for related-key differential
  characteristics in byte-oriented block ciphers: Application to aes, camellia,
  khazad and others.
\newblock {\em Annual International Conference on the Theory and Applications
  of Cryptographic Techniques},  pp. 322--344. Springer.

\bibitem{he2016deep}
He, K., Zhang, X., Ren, S., and Sun, J. Deep residual learning for image
  recognition.
\newblock {\em Proceedings of the IEEE conference on computer vision and
  pattern recognition},  pp. 770--778.

\bibitem{hu2018squeeze}
Hu, J., Shen, L., and Sun, G. Squeeze-and-excitation networks.
\newblock {\em Proceedings of the IEEE conference on computer vision and
  pattern recognition},  pp. 7132--7141.

\bibitem{chen2021new}
Chen, Y., Shen, Y., Yu, H., and Yuan, S.
\newblock A new neural distinguisher considering features derived from multiple
  ciphertext pairs.
\newblock bxac019.

\bibitem{hou2021improve}
Hou, Z., Ren, J., and Chen, S. Improve neural distinguishers of simon and
  speck.
\newblock {\em Security and Communication Networks}, {\bf  2021}.

\bibitem{kingma2014adam}
Kingma, D.~P. and Ba, J.
\newblock Adam: A method for stochastic optimization.

\bibitem{lu2021effect}
Lu, J., Liu, Y., Ashur, T., and Li, C. On the effect of the key-expansion
  algorithm in simon-like ciphers.
\newblock {\em The Computer Journal}, {\bf  65}, 2454--2469.

\end{thebibliography}

\end{document}